\documentclass[journal, a4paper]{IEEEtran}
\usepackage{amsmath,graphicx}
\usepackage{xcolor}
\usepackage{hyperref}

\usepackage{dblfloatfix}

\usepackage{enumitem}
\setlist{nosep, leftmargin=14pt}

\usepackage{mwe} 

\usepackage{algorithm}
\usepackage{algorithmic}

\usepackage{amsmath,amsfonts,bm}

\def\1{\bm{1}}

\def\vr{{\bm{r}}}

\def\vv{{\bm{v}}}

\def\vy{{\bm{y}}}
\def\vz{{\bm{z}}}

\def\mF{{\bm{F}}}

\def\mP{{\bm{P}}}

\DeclareMathAlphabet{\mathsfit}{\encodingdefault}{\sfdefault}{m}{sl}
\SetMathAlphabet{\mathsfit}{bold}{\encodingdefault}{\sfdefault}{bx}{n}

\def\gE{{\mathcal{E}}}

\def\gL{{\mathcal{L}}}

\def\gP{{\mathcal{P}}}

\newcommand{\R}{\mathbb{R}}

\newcommand{\T}{\top}

\title{Memory-efficient optimization of implicit neural representations for CT reconstruction}

\author{Mahrokh Najaf, Gregory Ongie
\thanks{Marquette University, Department of Mathematical and Statistical Sciences, Milwaukee, WI, USA. Corresponding author: gregory.ongie@marquette.edu}}

\begin{document}
\maketitle

\begin{abstract}
Implicit neural representations (INRs) provide a parameter-efficient and fully differentiable image model for CT reconstruction. However, optimizing INRs for CT reconstruction using standard auto-differentiation techniques can be prohibitively GPU memory-intensive, especially in 3D imaging, due to the large number of INR evaluations needed to simulate ray projections. To address this issue, we propose a memory-efficient stochastic gradient approximation based on decomposing the gradient into a Jacobian–vector product that is amenable to stochastic subsampling. This approximation allows the user to trade-off between GPU memory usage and gradient approximation accuracy. Our experiments on synthetic 2D data demonstrate that gradient approximation uses far less GPU memory than standard INR training, while yielding reconstructions that are comparable in convergence behavior and mean squared error. Finally, we demonstrate that the proposed approach allows for memory-efficient 3D cone beam CT reconstruction in a sparse-view setting.
\end{abstract}



\section{Introduction}
Implicit neural representations (INRs) provide a parameter-efficient, nonlinear, differentiable model for image reconstruction tasks in CT. In an INR, the unknown attenuation image is represented by a small-scale neural network $f_\theta:\R^d\to\R$ that takes spatial coordinates as input and outputs the corresponding grayscale value. Fitting an INR to CT measurements can be viewed as an unsupervised, model-based iterative reconstruction (MBIR) approach, where the unknown is parameterized by the network weights $\theta$ rather than by pixels/voxels. Recent work has explored INRs for low-dose and sparse-view CT reconstruction \cite{sun2021coil, zha2022naf, shen2022nerp, fu2024attenuation}, demonstrating that INRs can yield high-quality reconstructions while implicitly regularizing against noise and artifacts.

Given measured sinogram data,  INR parameters are optimized to minimize a data-fidelity loss that enforces consistency with the forward CT model. While conceptually appealing, this approach introduces a significant computational challenge: training an INR typically requires differentiating through the CT forward operator. When auto-differentiation is used end-to-end, the computational graph must store intermediate states associated with both the neural network and the projection operator. As a result, peak GPU memory consumption grows rapidly with image resolution, number of views, and dimensionality. These memory constraints force practitioners to either use very small INRs with limited representational capacity, or to severely restrict the number of ray projections simulated per iteration. In 3D cone-beam CT, this limitation is particularly acute: computing and storing the gradients associated with even a single view can exceed available GPU memory on typical hardware (e.g., reconstruction of a $501^3$ voxel volume on a 24 GB GPU; see Section~\ref{sec:exp:3d}). This motivates the need for alternative optimization strategies that reduce memory usage while preserving reconstruction quality.

In this paper, we propose a memory-efficient gradient approximation strategy for training INRs that avoids backpropagation through the CT forward model. Rather than relying on auto-differentiation to compute gradients of the loss with respect to the INR parameters, we explicitly form a Jacobian–vector product that separates the projection operator from the network derivative. This allows the use of efficient projection and backprojection routines, combined with stochastic coordinate subsampling, to approximate the gradient with substantially reduced peak GPU memory requirements.

We illustrate the proposed approach for sparse-view CT reconstruction on synthetic 2D data and real 3D data with a variety of INR architectures. Our results show that the proposed gradient approximation requires far less peak GPU memory than standard auto-differentiation, and the resulting reconstructions are comparable to auto-differentiation in terms of convergence behavior and final image mean squared error. 

\section{Problem formulation}
Let $\vy \in \mathbb{R}^m$ denote measured CT data (a sinogram in 2D or cone-beam projections in 3D), rearranged as a vector. Let $f_\theta:\mathbb{R}^d\to\mathbb{R}$ denote an INR with parameters $\theta\in\mathbb{R}^p$ that maps $d$-dimensional ($d=2,3$) spatial coordinates to attenuation values. Also, let $\gP$ denote the forward projection operator that approximates ray integrals of a continuously defined attenuation function.

INR-based CT reconstruction is posed as the optimization problem:
\begin{equation}\label{eq:inrfit}
\min_{\theta} \gL(\gP\{f_\theta\},\vy),
\end{equation}
where $\gL:\R^{m}\times \R^m\rightarrow \R$ is a given loss function. In this work, we focus on two loss functions: the least squares (LS) loss
$\mathcal{L}_{\text{LS}}(\vz, \vy) = \tfrac{1}{2}\|\vz - \vy\|^2$
and the filtered least squares (FLS) loss
$\mathcal{L}_{\text{FLS}}(\vz, \vy) = \tfrac{1}{2}(\vz - \vy)^\top\mF(\vz-\vy),$
where $\mF \in \R^{m\times m}$ is a matrix that represents convolution with a ramp filter along detector columns, as used in filtered back-projection. The FLS loss can be viewed as a preconditioned least squares objective that improves conditioning of the optimization problem. This loss was investigated for INR reconstruction in our prior work \cite{najaf2025accelerated}. While we focus on the LS and FLS loss in our experiments, we  emphasize that the gradient approximation approach proposed below applies to \emph{any} loss $\gL(\cdot,\cdot)$ that is differentiable in its first argument.

The most direct approach to minimizing \eqref{eq:inrfit} is to use auto-differentiation to obtain gradients of the loss, enabling the use of first-order optimizers such as Adam. However, auto-differentiation requires retaining the computational graph (and saved intermediate outputs) for the forward projector so that gradients can be computed during the backward pass. This can consume substantial GPU memory, and the cost is amplified in 3D due to the larger volume size and the increased number of projection rays.

\section{Proposed Gradient Approximation}

We now describe our proposed memory-efficient approximation to gradients of \eqref{eq:inrfit}. First, we decompose $\gP(f_\theta) = \mP\gE\{f_\theta\}$, where $\gE\{f_\theta\} = [f_\theta(x_1),...,f_\theta(x_n)]^\T\in\mathbb{R}^n$ denotes the evaluation operator that forms an image/volume by sampling $f_\theta$ on a regular grid of coordinates $x_1,...,x_n \in \R^d$ within a field-of-view (FOV) mask, and $\mP\in\mathbb{R}^{m\times n}$ denotes a discrete forward projection operator.
Let $L(\theta) = \gL(\vz,\vy)$ be the loss function, where $\vz = \gP(f_\theta) = \mP\gE\{f_\theta\}$ is implicitly a function of INR parameters $\theta$, and let $\vr^\T = \partial_{\vz} \gL(\vz,\vy)$. Then by the Jacobian chain rule we have
\[
\nabla L(\theta)^\T = \frac{\partial L}{\partial \vz} \frac{\partial \vz}{\partial  \theta} = \vr^\T \mP \left[\frac{\partial}{\partial \theta}\gE\{f_\theta\}\right].
\]
Defining $\vv = \mP^\T \vr$, the gradient can be written as
\[
\nabla L(\theta) = \left[\frac{\partial}{\partial \theta}\gE\{f_\theta\}\right]^\T \vv = \sum_{i=1}^n v_i \nabla_\theta f_\theta(x_i).
\]
Our key observation is that the sum structure in the gradient above enables stochastic approximation via subsampling. Let $I \subset [n]$ denote a random subset of INR evaluation indices (e.g., random coordinate batches within the FOV mask). We approximate the gradient by
\begin{equation}\label{eq:grad_subsample}
\widehat{\nabla L}(\theta) = \frac{n}{|I|} \sum_{i\in I} v_i \nabla_\theta f_\theta(x_i).
\end{equation}
If the index sets $I$ are drawn uniformly at random, this is an unbiased estimator:
\[
\mathbb{E}_I\bigl[\widehat{\nabla L}(\theta)\bigr] = \nabla L(\theta).
\]
For implementation purposes, observe that if $\vv$ is treated as constant with respect to $\theta$, then $\widehat{\nabla L}(\theta) = \nabla_\theta \tilde{L}(\theta)$ where
\begin{equation}\label{eq:virtual_loss}
\tilde{L}(\theta) = \frac{n}{|I|} \sum_{i\in I} v_i f_\theta(x_i)
\end{equation}
is a ``virtual'' loss whose gradient matches the desired stochastic gradient estimate. Therefore, auto-differentiation can be applied to the \eqref{eq:virtual_loss} to obtain the gradient approximation $\widehat{\nabla L}(\theta)$. Algorithm \ref{alg} summarizes this approach.

\begin{algorithm}
\caption{Stochastic Gradient Estimation for INR-based CT Reconstruction}
\begin{algorithmic}[1]\label{alg}
\REQUIRE INR parameters $\theta$, measurements $\vy$, batch size $|I|$
\STATE Evaluate $\vz = \mP\gE\{f_\theta\}$ \hfill \COMMENT{no gradient tracking}
\STATE Compute $\vv = \mP^\T \partial_{\vz}\gL(\vz,\vy)$ \hfill \COMMENT{no gradient tracking}
\STATE Sample random index set $I \subset [n]$, $|I| \ll n$
\STATE Compute virtual loss: $\tilde{L} = \frac{n}{|I|} \sum_{i \in I} v_i f_\theta(x_i)$
\STATE \textbf{return} $\nabla_\theta \tilde{L}$ via auto-diff. \hfill \COMMENT{gradient tracking on $\theta$ only} 
\end{algorithmic}
\end{algorithm}

Note that for the LS loss, $\partial_{\vz}\gL_{\text{LS}}(\vz,\vy) = (\vz-\vy)$, and for the FLS loss $\partial_{\vz}\gL_{\text{FLS}}(\vz,\vy) = \mF(\vz-\vy)$. Therefore, Step 2 in Algorithm \ref{alg} reduces to backprojection of the residual $\vz-\vy$ for the LS loss, and a filtered backprojection of the residual for the FLS loss, which can be efficiently computed outside the computational graph.

Finally, note that the proposed gradient approximation in Algorithm \ref{alg} is also compatible with ordered subsets-like approaches where the gradient of the loss is evaluated with respect to a partial subset of views.

\subsection{Memory and computational considerations}
The memory-requirements of computing gradients of the loss in \eqref{eq:inrfit} using auto-differentiation scales proportionally to the total number of INR evaluations used to estimate ray projections. For example, to estimate the projections associated with one view may require millions of INR evaluations, which can lead to large peak GPU memory usage. In contrast, the gradient approximation strategy presented computes the weight vector $\vv$ with optimized CT operators without gradient tracking, and backpropagates only through the INR evaluated on a coordinate minibatch. Consequently, peak GPU memory usage is dominated by the memory requirements to evaluate the INR over the minibatch, rather than the memory requirements to compute ray projections.

\begin{table}[!h]
\centering
\begin{tabular}{l|c|c|c}
\hline
 & FFNs & SIREN & Hash Enc. \\
\hline
Auto-differentiation  & 2.625 & 3.453 & 0.869 \\
Proposed gradient approx.& 0.708 & 0.767 & 0.534 \\
\hline
\end{tabular}

\vspace{4pt}

\caption{Peak GPU memory usage (in GB) across different INR architectures for 2D  reconstructions}
\vspace{-1em}
\label{tab:memory}
\end{table}
Table~\ref{tab:memory} illustrates the peak GPU memory usage for auto-differentiation versus our proposed gradient approximation across INR architectures for our 2D reconstruction experiments (see Section \ref{sec:exp:2d} for more details). Observe that the gradient approximation requires far less peak GPU memory than auto-differentiation, allowing for substantially larger problem sizes.


\begin{figure*}[!tp]
\centering
\includegraphics[width=0.72\linewidth]{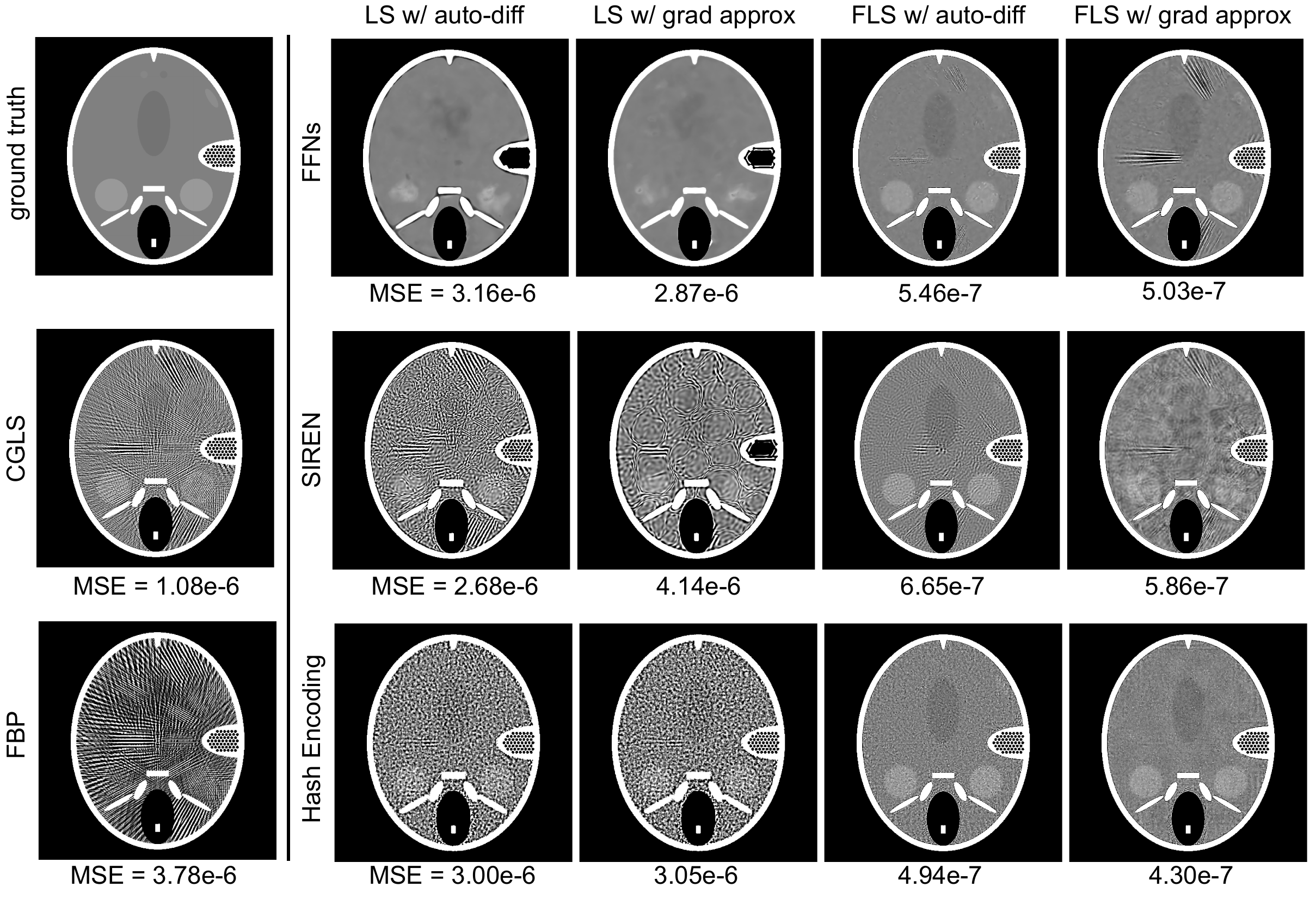}
    \caption{\small 2D FORBILD sparse-view reconstructions obtained with conjugate gradient least squares (CGLS) and filtered backprojection (FBP), alongside reconstructions produced by training an INR using FFNs, SIREN, and Hash Encoding architectures. All images are shown on the scale [0.0200,0.0220] mm$^{-1}$.}
    \label{fig:forbild}
\end{figure*}

\begin{figure*}[!htb]
\centering
\includegraphics[width=0.73\linewidth]{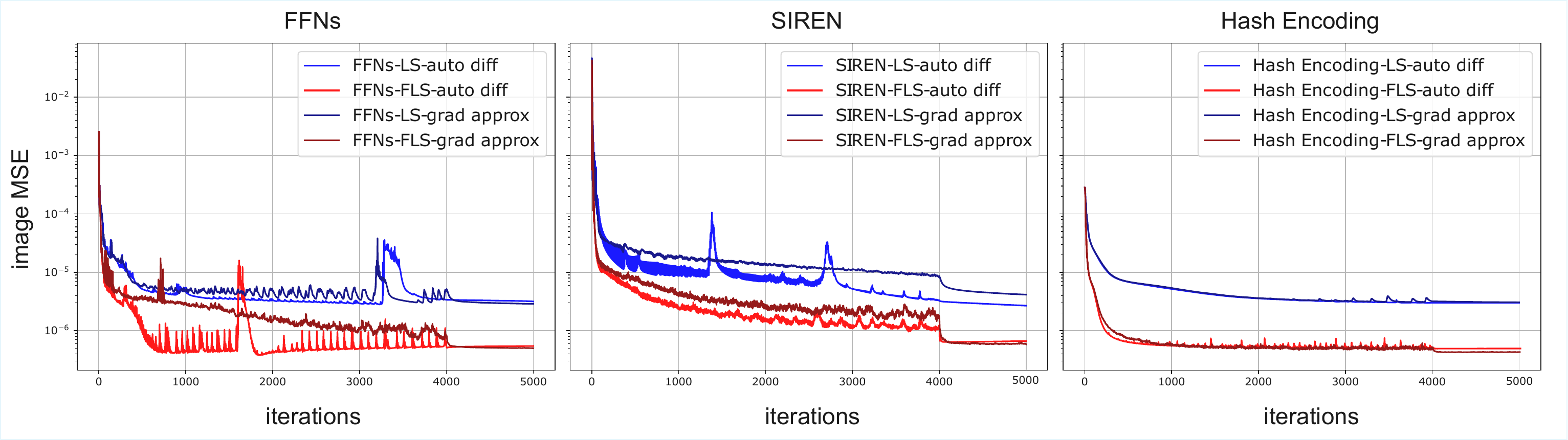}

\vspace{0.6em}

\includegraphics[width=0.72\linewidth]{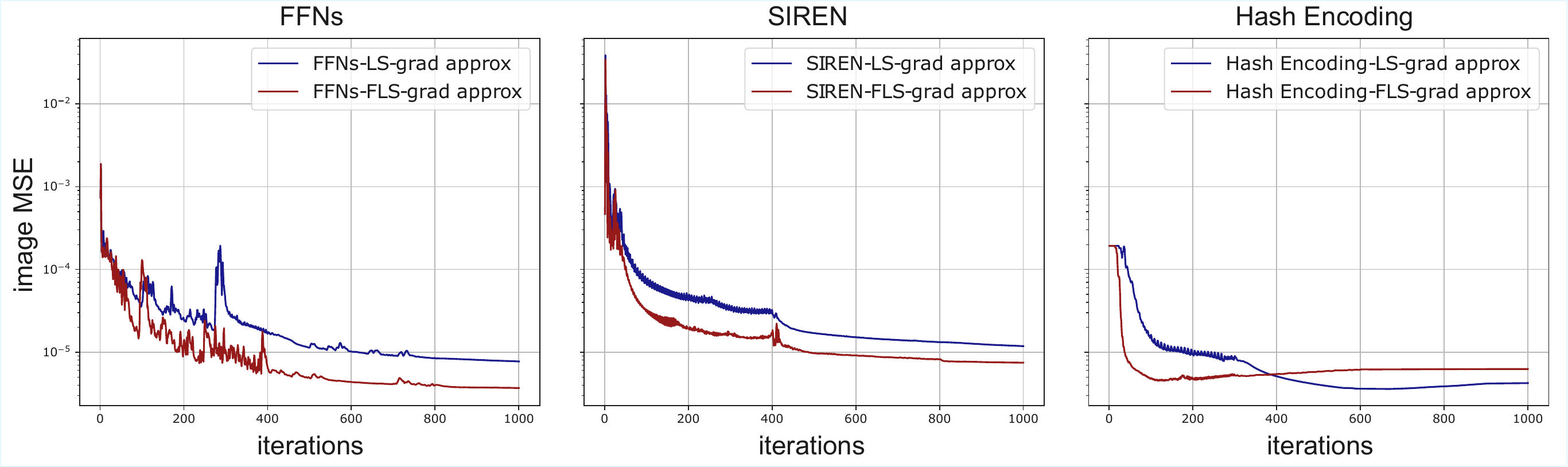}

\caption{\small Top row (2D FORBILD): Image MSE plots for INR training with FFNs, SIREN, and hash encoding, comparing auto-differentiation and gradient approximation (for both LS and FLS).
Bottom row (3D Walnut): Image MSE plots for INR training with FFNs, SIREN, and hash encoding using gradient approximation (for both LS and FLS).}
\label{fig:mse_plots}
\end{figure*}

\section{Experiments}\label{sec:exp}
Here we demonstrate the proposed approach for sparse-view CT reconstruction. We present results for three popular INR architectures: fully-connected ReLU networks with an initial Fourier features encoding (FFNs) \cite{tancik2020fourier}, fully-connected networks with sinusoidal activation functions (SIREN) \cite{sitzmann2020implicit}, and a MLP with an initial multi-resolution hash encoding (Hash Enc.) \cite{muller2022instant}.

We use the LivermorE AI Projector for Computed Tomography (LEAP) software package\footnote{\url{https://github.com/llnl/LEAP}} for GPU-enabled forward and backward projections and FBP reconstructions \cite{kim2023differentiable}. All experiments were conducted using a single NVIDIA GeForce RTX 4090 GPU with 24 GB RAM. 

As baselines, we compute reconstructions using analytical methods, i.e., filtered backprojection for 2D fan-beam and FDK for 3D cone-beam, both referred to as ``FBP'' in this paper. We also compare against a pixel/voxel-space least squares reconstruction obtained via the conjugate gradients algorithm (CGLS).

\subsection{2D FORBILD Phantom}\label{sec:exp:2d}
In our first set of experiments, we synthetically generate a sparse-view 2D fanbeam acquisition of the 2D FORBILD phantom (180 equi-spaced views along a 360 degree arc, with 512 detector pixels). To avoid the inverse crime, we generate projection data from the phantom rasterized to a $2048\times2048$ pixel grid, and for reconstruction use projection operators acting on a $512\times 512$ pixel grid.  

 We compare four training configurations for each INR architecture: (1) LS loss with auto-differentiation (LS w/auto-diff), (2) LS loss with the proposed gradient approximation (LS w/grad-approx), (3) FLS loss with auto-differentiation (FLS w/auto-diff), (4) FLS loss with the proposed gradient approximation (FLS w/grad-approx). For the gradient approximation cases, we use a minibatch size of $n/16$, where $n$ is the number of pixel locations inside the FOV.

Figure \ref{fig:forbild} shows reconstructions obtained after training an INR for 5000 iterations with the Adam optimizer, which are comparable to those produced by CGLS or FBP, while in some cases achieving higher quality. The top panel of Figure \ref{fig:mse_plots} plots the pixel-wise mean-squared-error (MSE) of the INR reconstructions versus the ground truth phantom over all iterations. Note that for a given loss function and INR architecture, we see similar convergence behavior with and without the proposed gradient approximation approach.

\vspace{-0.5em}

\subsection{3D Walnut Dataset}\label{sec:exp:3d}

We also illustrate the scalability of our approach by reconstructing real cone-beam CT data of a walnut \cite{der2019cone}. The original dataset contains 1201 projection views with $972\times 768$ detector pixels, which we retrospectively subsample to 120 views. The reconstructed volume size is $501^3$ voxels. We use a CGLS reconstruction obtained from the full set of 1201 views as the ground truth.

For this dataset, auto-differentiation using the full set of 120 views was not feasible due to GPU memory constraints. Therefore, we report results only for the proposed memory-efficient gradient approximation approach. We used a minibatch size of $n/512$, where $n$ is the number of voxel locations inside the FOV. This resulted in the following peak GPU memory usages: 7.96 GB for FFNs, 8.43 GB for SIREN, and 6.36 GB for Hash Encoding.

\begin{figure}[!htb]
\includegraphics[width=\columnwidth]{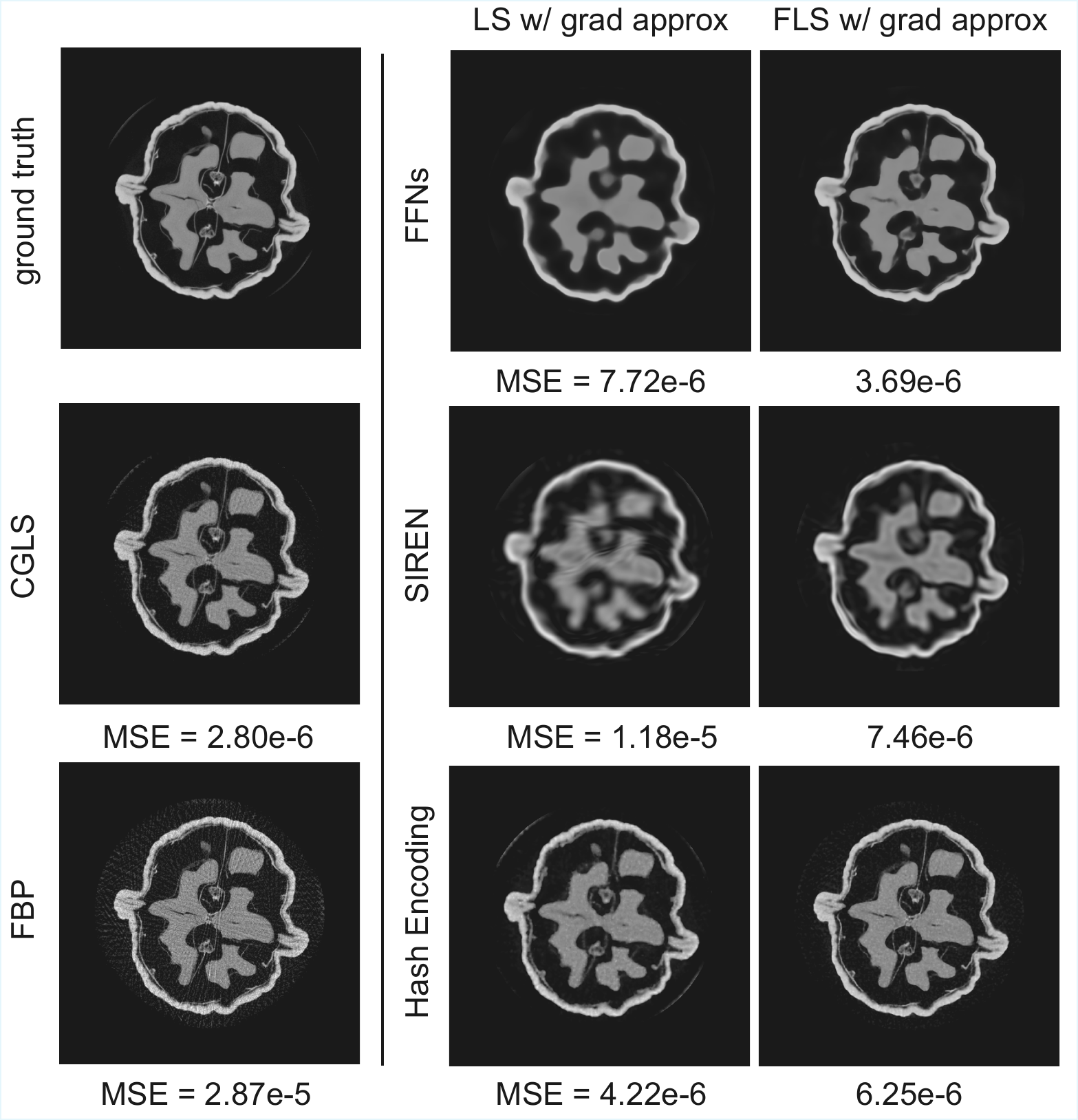}
    \vspace{-1.5em}
    \caption{\small 3D sparse-view cone-beam CT reconstructions obtained with conjugate gradient least squares (CGLS) and filtered backprojection (FBP), alongside reconstructions produced by training an INR using FFNs, SIREN, and Hash Encoding architectures. All images are shown on the scale [0.00,0.08] mm$^{-1}$.}
    \label{fig:walnut}
\end{figure}

Figure \ref{fig:walnut} shows reconstructions obtained after training an INR for 1000 iterations using the Adam optimizer. All three INR architectures (FFNs, SIREN, and Hash Encoding) produce reconstructions that are comparable to those from CGLS and FBP. The bottom panel of Figure \ref{fig:mse_plots} plots the pixel-wise MSE of the INR reconstructions versus the ground truth phantom over all iterations, which shows stable convergence of the proposed approach.


\section{Conclusion}
We introduced a GPU memory-efficient gradient approximation strategy for optimizing INRs in CT reconstruction. Our experiments show that the proposed gradient approximation dramatically reduces peak GPU memory while producing reconstructions that are comparable to auto-differentiation in convergence behavior and final MSE. Finally, we also demonstrate that the proposed approach allows for easy scalability to 3D cone beam reconstruction, where the GPU memory demands of auto-differentiation through the CT projector can be prohibitive.

\vspace{-1em}

\bibliographystyle{IEEEbib}
\bibliography{strings,refs_new}

\end{document}